\documentclass[aps, prb, reprint, superscriptaddress, footinbib]{revtex4-2}

\newcommand{\diff}{\mathop{}\!d}
\usepackage{amssymb}

\usepackage[utf8]{inputenc}

\usepackage{circuitikz}
\usepackage{tikz,tkz-euclide}
\usepackage{tikz-3dplot}

\usepackage{graphicx}

\usepackage{xcolor}

\usepackage{subfigure}

\definecolor{flored}{HTML}{D7191C}
\definecolor{floorange}{HTML}{FDAE61}
\definecolor{floyellow}{HTML}{FFFFBF}
\definecolor{flolightgreen}{HTML}{A6D96A}
\definecolor{flodarkgreen}{HTML}{1A9641}

\usepackage{amsmath}

\usepackage{hyperref}
\usepackage{url}

\usepackage[capitalise]{cleveref} 

\begin{document}
	
	
	\author{Florian St{\"a}bler}
		\affiliation{Département de Physique Théorique, Université de Genève, CH-1211 Genève 4, Switzerland}
	\author{Eugene Sukhorukov}
		\affiliation{Département de Physique Théorique, Université de Genève, CH-1211 Genève 4, Switzerland}
	
	
	\title{Transmission line approach to transport of heat in chiral systems with dissipation}
	\date{\today}

\begin{abstract} Measurements of the energy relaxation in the integer quantum hall edge at filling factor \(\nu=2\)  suggest  the breakdown of heat current quantization [H. le Sueur et al., Phys. Rev. Lett. 105, 056803]. It was shown, in a hydrodynamic model, that dissipative neutral modes contributing apparently less than a quantum of heat can be an explanation for the missing heat flux [A Goremykina et al., arXiv preprint arXiv:1908.01213]. This hydrodynamic model relies on the introduction of an artificial high-energy cut-off and lacks a way of a priori obtaining the correct definition of the heat flux.  In this work we overcome these limitations and present a formalism, effectively modeling dissipation in the quantum hall edge, proving the quantization of heat flux for all modes. We mapped the QHE to a transmission line by analogy and used the Langevin equations and scattering theory to extract the heat current in the presence of dissipation. 
\end{abstract}

\maketitle

\section{Introduction}

The missing heat paradox, reported in the integer quantum hall edge at filling factor \(\nu=2\) \cite{le_sueur_energy_2010,granger_observation_2009,venkatachalam_local_2012}, challenges two of the most fundamental phenomena in mesoscopic and nanoscopic physics. On one hand many experiments confirm that heat flux is quantized, more specifically, the rate at which any type of carrier can transport heat at most in a ballistic channel is proportional to a universal value known as the heat flux quantum quantum  \(J_q= \frac{\pi k_B^2}{12 \hbar}T^2\) \cite{pendry_quantum_1983,kane_quantized_1997}. Ballistic channels can be found in a variety of different systems including quasi-1D semiconductor nanostructures \cite{schwab_measurement_2000}, carbon nanotubes \cite{brown_ballistic_2005} or Quantum Hall (QH) systems, both at integer \cite{le_sueur_energy_2010,granger_observation_2009} and fractional fillings \cite{banerjee_observation_2018,banerjee_observed_2017,mross_theory_2018}. On the other hand introducing dissipation may resolve the paradox, since it is suggested to be able  to break the aforementioned heat flux quantization\cite{goremykina_heat_2019}.

In contrast to charge transport which is often protected by symmetries or topology \cite{hasan_colloquium_2010}, heat transport in such systems requires a more elaborate theoretical framework. Taking into account the smooth confining potential and screening of the gate electrodes in QH systems, edge states manifest themselves as a charge density profile consisting of alternating compressible and incompressible strips \cite{chklovskii_electrostatics_1992,chamon_sharp_1994}. This picture has been experimentally confirmed \cite{pascher_imaging_2014}.  The effects of interaction, disorder or finite temperature effects predict additional non-trivial neutral counter propagating excitations in the edge \cite{meir_composite_1994,wang_edge_2013,yang_field_2003,joglekar_edge_2003,zhang_theoretical_2014,protopopov_transport_2017,park_topological_2015,wan_reconstruction_2002,ferconi_edge_1995}. Aleiner and Glazman analyzed the low-energy spectrum of excitations of a compressible electron liquid in a strong magnetic field and showed that the integer QH edge can host neutral co-propagating (AG) excitation \cite{aleiner_novel_1994}. Theory predicts an infinite number of neutral downstream AG excitations, however they were never detected in experiment. What has been measured is a leakage of the injected energy into the QH edge at different integer fillings \cite{granger_observation_2009,venkatachalam_local_2012}, suggesting the presence of additional degrees of freedoms for energy to be redistributed.

A detailed study of the QH edge at filling factor \(\nu=2\) followed \cite{le_sueur_energy_2010}. In the experiment, energy was injected in the form of Joule heat into a QH edge at a constant rate. This creates a non-equilibrium distribution function which eventually relaxes to a ``hot"  Fermi distribution function, which can be probed by a quantum dot downstream. They found no energy transfer towards the excitation of thermalized states and an efficient energy redistribution between the two channels without particle exchanges. However, after long equilibration lengths \(L \geq 10\mu m\) the corresponding temperature of the equilibrium distribution function saturated at a value which was \(13\% \) lower than expected for two interacting edge channels. The effective temperature becomes independent of the propagation length indicating, that no energy leaked into the bulk of the system or is lost to some external mechanisms not taken into account by the experimental procedure. The authors concluded that the presence of additional degrees of freedom can explain the outcome of the experiment, however, to match the numbers this additional mode would need to carry less than a quantum of heat to explain the loss.

 \begin{figure}[htbp] 
	\centering
	\includegraphics{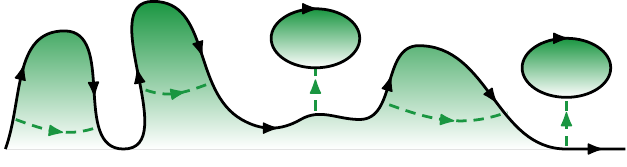}
	\caption{Possible origin of dissipation in the QH edge: Due to imperfections of the edge, electrons have the possibility to tunnel, resembling a simple reservoir for electrons. This allows us to make an analogy with a transmission line, which has distributed capacitors and resistors.  The reservoirs act as a heat bath with a self consistently determined temperature.}
	\label{fig:origindis}
\end{figure}

 In the previous work we addressed the possibility of a third AG excitation carrying less than a heat flux quantum due to dissipation \cite{goremykina_heat_2019}. The charged mode of the system carries a flux quantum \(J=J_q=\frac{\pi}{12 \hbar \beta^2}\). The neutral mode subject to a transverse current proportional to the longitudinal conductivity \(\sigma_{xx}\) in the  compressible strip will result in this mode having a dissipative term in its low-energy spectrum and thus carrying apparently a  reduced amount of heat. The model is limited to the low-energy degrees of freedoms with a wavelength much larger than the inverse size of the compressible strip due to the inhomogenity of the edge in the transverse direction. Despite the introduction of this artificial cut-off the found reduction of the carried heat is universal \(J=\frac{\sigma_{xx}}{2\pi\sigma_{xy}}J_q\).  We also note here, that the reported loss of heat might be due to non-local relaxation mechanisms \cite{rodriguez_relaxation_2020,lunde_statistical_2016,fischer_interaction-induced_2019,krahenmann_auger-spectroscopy_2019}. However, the results of the present paper show a more complete picture of the low energy theory of the edge.

\textit{The goal of this paper} is to apply a combination of Langevin equations and scattering states to model dissipation in chiral systems effectively \cite{sukhorukov_scattering_2016}. We attach a chiral system to a bath modeled as an open system, having the advantage to  be able to address equilibrium and non-equilibrium situations,  as well as being analytically treatable. These advantages make it a new approach to modeling dissipation, complementary to the  Caldeira-Leggett model, which successfully captured the features of dissipation on a quantum level. We provide an effective theory for chiral dissipative systems which is not restricted to the low-energy degrees of freedom and prove the quantization of heat for these systems, which, to the best of our knowledge, has never been addressed. Furthermore, we present the correct definition of heat flux, which is unobtainable starting from a hydrodynamic point of view, we try to resolve the experimental paradox of the 'missing heat' and discuss the role of AG modes in QH systems.

\section{Theoretical Model}

Our goal is to capture the physics of a compressible strip in the presence of disorder. We propose a minimal model and focus on the experimental situation of filling factor \(\nu=2\) and model two co-propagating modes with a typically large (spin-) resistance between them  \footnote{Note that we consider in the present paper a model at filling factor \(\nu=2\) and explicitly model a (spin-flip) resistance between the two modes. However this is completely analogue to modelling the hydrodynamic edge reconstruction, where the additional mode is only half filled. Our results can be simply connected to the hydrondynamic model in \cite{goremykina_heat_2019} by the transformation \(R_q\rightarrow R_q/2\).}. Interactions and dissipation can be conveniently introduced using a transmission line approach. We formally discretize the  system into many nodes which interact longitudinally (within the same mode) with a chiral quantum resitor of strength \(R_q= \frac{2\pi \hbar}{e^2}\) and transversely (between the modes) with a quantum resistor of, in principle, arbitrary
 strength \(R_\perp\).

\begin{figure}[htbp]%
	\centering
	\subfigure[Transmission line]{%
		\label{fig:TLa}%
		\includegraphics[width=.48\linewidth]{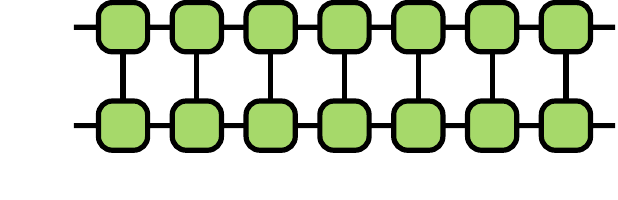}}%
	\quad
	\subfigure[Single node]{%
		\label{fig:TLb}%
		\includegraphics[width=.48\linewidth]{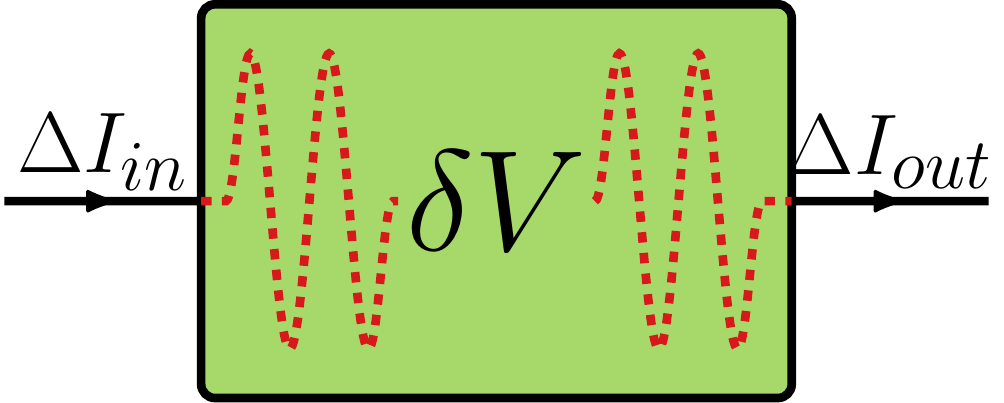}}%
	\caption{ \cref{fig:TLa} shows the transmission line. The two copropagating modes consist of many reservoirs, which can interact longitudinally and transversely. \cref{fig:TLb} shows a single node. The incoming current is fully dissipated in the ohmic reservoir heating it in return. The outgoing current contains a contribution of the collective mode \(\delta V(t)/R_q\) and a Langevin source contribution \(\delta I(t)\) due to the thermal noise of the resistor.}
\end{figure}

 One can view this discretization also as the attempt to model inhomogenities of the edge, see \cref{fig:origindis}. Electron's moving along the edge might be disturbed or stored in these inhomogenities for some time and then are reemitted at a later time. This physically resembles an Ohmic reservoir similar to the one presented in \cite{slobodeniuk_equilibration_2013}. We consider the situation where the level spacing of the reservoir is smaller than the charging energy, the other limit will be considered elsewhere. Thermal fluctuations incident to the ohmic contact are absorbed and lead to the creation of  voltage \(\delta V(t)\) and  current fluctuations \(\delta I(t)\) being reemitted. This can be captured within the framework of Langevin equations \footnote{Note that the same Langevin equations follow directly from solving the equation of motion for two capacitively interacting chiral bosons within the ohmic contact and assigning the correct boundary currents to the solution}.

\begin{gather}
	\frac{\diff}{\diff t} \delta Q(t)= \Delta I_{\text{in}}(t)-\Delta I_{\text{out}}(t),\\
	\Delta I_{\text{out}}= \frac{\delta Q(t)}{R_q C} +\delta I(t).
\end{gather} where the first equation is Kirchoff's law guaranteeing current conservation in the Ohmic contact and the second equation is the Langevin equation with \(C\) being the capacitance of the node. In our notation $\Delta I$ corresponds to the total fluctuation of current containing charge and current fluctuations and $\delta I$ refers to the current fluctuations in the form of a Langevin source. We take the Langevin equation as a starting point and introduce two chains of \(N\) nodes interacting with a transverse resistor. The equation of motion for the voltage fluctuation \(\delta V^\alpha_j=\delta Q^\alpha_j/C\) at position \(j\) in the upper/lower part of the TL $\alpha=1,2$ is given by

\begin{gather}\label{eq:eom1}
	\frac{\diff }{\diff t} \delta Q^\alpha_j = \Delta I^\alpha_{\text{in},j}-\Delta I^\alpha_{\text{out},j} \mp \Delta I^\perp_{j},\\\label{eq:eom2}
	\Delta I^\perp_{j} =  \frac{1}{R_\perp }\left( \delta V^{(1)}_{j}-\delta V^{(2)}_{j}\right) +  \delta I_{j}^{\perp},
	\\ \Delta I^\alpha_{\text{out},j} = \frac{1}{R_q} \delta V^\alpha_{j} +  \delta I_{j}^{\alpha} \label{eq:eom3}
\end{gather} where the incoming current \( \Delta I^\alpha_{\text{in},j}=\Delta I^{\alpha}_{\text{out},j-1}  \) \footnote{Adding retardation between the nodes does not change our analysis. The bosonic currents are related through plane wave propagation \( \Delta I^\alpha_{\text{in},j}=\Delta I^{\alpha}_{\text{out},j-1} e^{i \frac{\omega}{v_F}\xi}  \), where \(\xi\) is the distance between the nodes. This phase can be removed from all integrals by a constant shift of the momentum integral. We thus neglect the phase shift here. } is given by  the outgoing current of the previous node, $\delta I_j^\alpha$ and $\delta I_j^\perp$ denote a Langevin source and the negative (positive) sign is chosen for  \(\alpha=1 (2)\).

\begin{figure}[htbp] 
	\centering
	\includegraphics[width=0.4\linewidth]{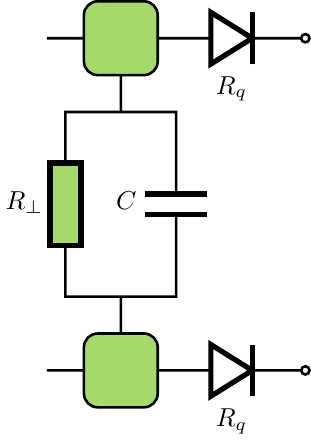}
	\caption{This figure shows a transverse cross section of the transmission line depicted in  \cref{fig:TLa}. The ohmic reservoirs longitudinally emit a current according to Ohm's law with the resitance \(R_q\) plus thermal fluctuations given by \cref{eq:eom3}. Furthermore the upper and lower mode interact via a transverse resistor \(R_\perp\) according to \cref{eq:eom2}. The capacitor represents the self capacitance \(C\) of the ohmic contacts.   }
	\label{fig:chiralequivcirc}
\end{figure}

The outgoing current fluctuations consist of the collective mode contribution and thermal fluctuations. We define the symmetric and anti symmetric combination of all voltages and currents, i.e. the charged and neutral mode \(X^{c/n}=\frac{1}{2}\left( X^{(1)}\pm X^{(2)}\right)\).

\begin{gather}\label{eq:VC}
	\frac{d }{dt} \delta Q^c_j \! =  \Delta I^{c}_{\text{out},j-1}\!\!-\Delta I^{c}_{\text{out},j} ,\\ \label{eq:VN}
	\frac{d }{dt} \delta Q^n_j \! =  \Delta I^{n}_{\text{out},j-1}\!\!-\Delta I^{n}_{\text{out},j} \!-\! \frac{2}{R_\perp} \delta V_j^n-\!\delta I_{j}^{\perp}.
\end{gather} The following Fourier transformation
\begin{gather} 
	X_j(t)=\sum\limits_{k=0}^{N-1} \int \frac{\diff \omega}{2\pi} e^{i\frac{2 \pi k}{N} j-i\omega t} X_k(\omega),\\ 	X_k(\omega)=\frac{1}{N}\sum\limits_{j=0}^{N-1} \int \diff t  e^{-i \frac{2 \pi k}{N}  j+i \omega t} X_j(t),
\end{gather}
in time and position allow us to formally solve the equation of motions and express the collective mode contribution \(\delta V^{c/n}\) as a function of the Langevin sources \(\delta I^{c/n}\) and \(\delta I^\perp\), which have a known correlation function. Furthermore, we can read off the spectra of the charged mode

\begin{equation}\label{eq:dispc}
  -i \omega_k^c=\left(R_q C\right)^{-1}\left(e^{-2 \pi i k /N}-1\right),
\end{equation}

 and neutral mode

 \begin{equation}\label{eq:dispn}
 	-i \omega_k^n=\left(R_q C\right)^{-1}\left(e^{-2 \pi i k /N}-1-\Gamma\right),
 \end{equation}
 which contain highly nonlinear terms responsible for dissipation and dispersion of the modes. The neutral mode contains furthermore an additional imaginary part governed by the strength of the perpendicular resistor \(\Gamma=2R_q/R_\perp\).

\subsection{Computation of the heat flux}

The advantage of the TL formulation is that we can compute the heat flux locally by considering a cross section in between nodes. Each node is connected to perfect chiral channels, a Hamiltonian system, for which a continuity equation for the energy density  \(\partial_t \hat{h}+ \partial_x \hat{J}=0\) can be derived. Starting from the Hamiltonian density of free chiral bosons \(\hat{h}=\frac{\hbar v_F}{4\pi}  \left(\partial_x \phi(x,t) \right)^2\), applying the equation of motion \(\partial_t \phi(x,t)+v_F \partial_x \phi(x,t) = 0\) and using the definition for the bosonic charge density \(\hat{\rho} (x,t) = \frac{e}{2\pi} \partial_x \phi(x,t)\) and bosonic current \(\hat{j}(x,t)= -\frac{e}{2\pi}\partial_t \phi(x,t)\). We arrive at the following expression for the average heat flux \nocite{levkivskyi_energy_2012}\footnote{The measured heat flux can deviate from the actual heat flux computed here. Since the chiral modes connecting the ohmic nodes have a perfectly linear dispersion the measured and actual heatflux will coincide \cite{levkivskyi_energy_2012}. We will consider the situation where the two heat fluxes might deviate elsewhere.}

\begin{equation}\label{eq:hf1}
	J= \left\langle \hat{J}\right\rangle= \frac{\pi \hbar}{e^2} \left\langle \hat{j}(x,t)^2\right\rangle.
\end{equation}

By means of Fourier transformation we obtain the current-current correlation function \( \left\langle \Delta I^{c/n}_{out,k}(\omega)\Delta I^{c/n}_{out,q}(\omega') \right\rangle \) from the equation of motion \cref{eq:VC,eq:VN}. Using inverse Fourier transform of \cref{eq:hf1} we can directly substitute the correlation function and obtain the heat carried by the charged and neutral mode. The details of how to solve the equations of motion and how to compute the resulting integrals can be found in the supplemental material.

\subsection{Noisepower from FDT}

 \cref{eq:hf1} allows us to express the heat flux through the correlation function of the Langevin sources  appearing in  \cref{eq:VC,eq:VN}. Due to locality in space and time of the Langevin sources the following simplification arises

\begin{equation}
\left\langle \delta I^{x}_{k}(\omega)\delta I^{x'}_{q}(\omega') \right\rangle = 2\pi \delta(\omega+\omega') \delta_{x,x'}\frac{\delta_{k,-q}}{N} S_{x}(\omega), 
\end{equation} where \(x,x' \in \{c,n,\perp\} \) labels the sources belonging to the charged, neutral or perpendicular source. Additionally, we have a Kronecker delta function for the momenta and a Dirac delta function for the frequencies with their proper normalization.  The noise spectral density   \( S_{x} \) is evaluated assuming a local thermal equilibrium in the channel, thus the Fluctuation Dissipation theorem (FDT) applies. 

\begin{equation}\label{eq:fdt}
	S_x(\omega)= G_x \left(\frac{\hbar \omega}{1-e^{-\beta \hbar \omega}}-S_{\text{vac}}(\omega)\right),
\end{equation} where the vacuum noise \(S_{\text{vac}}(\omega)=\hbar \omega \theta(\omega)\) is subtracted, with the inverse temperature \(\beta=\left(k_B T\right)^{-1}\), \(G_{c/n}= 1/(2 R_q)\) and \(G_\perp=\Gamma/R_q\). Note the difference of factor \(2\) between the longitudinal and transverse resistor, which is due to the chirality of the longitudinal resistors \cite{sukhorukov_scattering_2016}. With these ingredients the average heat flux can be computed. Note that for equilibrium noise, one finds the aforementioned heat flux quantum

\begin{equation}\label{eq:hfequil}
	J= \!  \int\limits_{-\infty}^\infty  \! \frac{\diff\omega}{4\pi} \left[\frac{\hbar \omega}{1-e^{-\beta \hbar \omega}}-\hbar \omega \theta(\omega)\right]=\frac{\pi}{12 \hbar \beta^2} = J_q.
\end{equation}

\section{Results}

The heat flux through the upper and lower arm can be expressed as the contribution carried by the charged mode and the neutral mode

\begin{equation}\label{eq:HF}
	\frac{J}{R_q} =\frac{1}{2} \sum_{\alpha=1,2} \left\langle \left(\Delta I^{\alpha}_{\text{out},j}\right)^2  \right\rangle =\! \sum_{\alpha'=c,n} \left\langle \left(\Delta I^{\alpha}_{\text{out},j}\right)^2  \right\rangle .
\end{equation}

We will compute the charged \(J^c\)  and neutral mode \(J^n\)  contributions separately. Note that the charged mode contribution can be computed simply by taking the limit \(J^c= \lim\limits_{\Gamma \rightarrow 0} J^n\). We proceed with the following steps: (i) take the limit of infinitely many nodes \(N\rightarrow \infty\), which allows us to replace the discrete \(k\)-summation by a \(k\)-integration, but keep the nodes separated by the distance \(\xi\). (ii) Compute the \(k\)-integral by mapping it onto the unit circle \(z\rightarrow e^{i k \xi}\) and use the residue theorem with the poles enclosed by the unit circle contour. (iii) Compute the \(\omega\) integral.

\subsection{Heat flux of the charged and neutral mode}

In this section we give the results for the fluxes computed as described above. For the charged mode we find

\begin{equation}\label{eq:hfc}
	J^c \! \!= \! \! \! \int\limits^{\infty\vphantom{\pi/\xi}}_{-\infty\vphantom{-\pi/\xi}} \int\limits^{\pi/\xi}_{-\pi/\xi} \! \!\frac{\diff\omega \diff k }{8\pi^2} \frac{\xi \omega^2 S_0(\omega)}{\left(\omega- \omega^c\left(k\right)\right)\left(\omega+ \omega^c\left(-k\right)\right)}\! = J_q,
\end{equation}  where the dispersion relation of the collective mode is given by \cref{eq:dispc} and we abbreviate the noise power with \(S_0(\omega)=\hbar \omega\left(1-e^{-\beta \hbar \omega}\right)^{-1}-S_{\text{vac}}(\omega)\). The charged mode carries a full flux quantum as expected. For the neutral mode we surprisingly find the following 

	\begin{equation}\label{eq:hfn}
	J^n \!\! =\hspace{-5pt} \int\! \! \frac{\diff\omega \diff k }{4\pi^2}\!    \frac{ \xi\left(R_q^2 C^2\omega^2 +\Gamma+\frac{\Gamma^2}{2}\right)S_0(\omega)}{R_q^2 C^2\left(\omega- \omega^n\left(k\right)\right)\left(\omega+ \omega^n\left(-k\right)\right)} \!= \! J_q,
\end{equation} with the limits of integration being the same as the ones for the charged mode and the spectrum of the neutral modes given by \cref{eq:dispn}. The heat carried by the neutral mode is given by a full flux quantum and completely independent of \(R_\perp\). This can be seen after computing the momentum integral, which makes it equivalent to \cref{eq:hfequil}; See the supplemental material for details on the calculation. This is unexpected since the dispersion relation of the neutral mode is subject to arbitrary strong dissipation, e.g. in the strongly damped limit \(\Gamma \rightarrow \infty\) . This is part of our proof, showing that heat is universally quantized even in the strongly overdamped regime.

\subsection{The sum rule}

 The second part of our proof can be understood as a generalized sum rule \cite{slobodeniuk_equilibration_2013} originating from each individual node of the transmission line. This is the direct manifestation of the unitarity of the scattering matrix at each node. This is one strength compared to the Caldeira-Leggett model our formalism offers.  One can straightforwardly solve \cref{eq:eom1,eq:eom2,eq:eom3} for the outgoing current \(\Delta I^{\alpha}_{\text{out},j}\) as a function of the incoming current \(\Delta I^{\alpha}_{\text{in},j}\) and the Langevin sources. This gives for the current fluctuations of the \(\alpha\in\{1,2\}\) channel
\begin{equation}
	\Delta I^{\alpha}_{\text{out},j} = \sum_{\beta= 1,2} \mathcal{T}^\alpha_{\text{in},\beta} \Delta I^{\beta}_{\text{in},j}+ \mathcal{T}^\alpha_{\parallel,\beta} \delta I^{\beta}_{j}+ \mathcal{T}^\alpha_{\perp} \delta I^{\perp}_{j}.
\end{equation} The corresponding coefficents are given by

\begin{gather*}
	\mathcal{T}^1_{\text{in},1} =\mathcal{T}^2_{\text{in},2} =\frac{1}{2}\left(A\left(\omega\right)+B\left(\omega\right)\right),\\
	\mathcal{T}^1_{\text{in},2}=\mathcal{T}^2_{\text{in},1} =-\mathcal{T}^1_{\parallel,2}=-\mathcal{T}^2_{\parallel,1}= \frac{1}{2}\left(A\left(\omega\right)-B\left(\omega\right)\right),\\
	\mathcal{T}^1_{\parallel,1}=\mathcal{T}^2_{\parallel,2}=1-\mathcal{T}^1_{\text{in},1},\\
	\mathcal{T}^1_{\perp}=-\mathcal{T}^2_{\perp}=-B\left(\omega\right),	
\end{gather*}with \(A(\omega)=\left(1-i R_q C \omega\right)^{-1}\) and \(B(\omega)=R_\perp \left(R_\perp +2 R_q -i R_\perp  R_q C \omega\right)^{-1}\). If one compute the noise power in the upper/lower channel one finds

\begin{multline}
	\left\langle \Delta I^{\alpha}_{\text{out},j}(\omega)\Delta I^{\alpha}_{\text{out},j}(-\omega) \right\rangle =\\= \sum_{\beta= 1,2} \left|\mathcal{T}^\alpha_{\text{in},\beta}\right|^2 \left\langle \Delta I^{\beta}_{\text{in},j}(\omega)\Delta I^{\beta}_{\text{in},j}(-\omega) \right\rangle+ \\ + \left|\mathcal{T}^\alpha_{\parallel,\beta}\right|^2 S_{\parallel,\beta} + \left|\mathcal{T}^\alpha_{\perp}\right|^2 S_{\perp}.
\end{multline} The sum rule states, that if the current incident to a node is equilibrium, i.e. it has a noise power given by \cref{eq:fdt} with \(G_{\text{in}} = \frac{1}{ R_q}\) we find that

\begin{equation}
	\sum_{\beta= 1,2} \left|\mathcal{T}^\alpha_{\text{in},\beta}\right|^2 G_{\text{in}} +\left|\mathcal{T}^\alpha_{\parallel,\beta}\right|^2 G_\parallel +\left|\mathcal{T}^\alpha_{\perp}\right|^2 G_\perp =\frac{1}{ R_q},
\end{equation} with \(G_{\text{in}}=G_\parallel=\frac{1}{R_q}\) and \(G_\perp = \frac{2}{R_\perp}\) .  This immediately explains the earlier findings since every outgoing current from a node will be equilibrium if the incident currents are equilibrium.

\subsection{Exact cancellation of Joule heating and backaction in equilibrium}

The structure of the outgoing current is always of the form \(\Delta I_{\text{out}}= R_q^{-1}\delta V(t) +\delta I(t)\), hence the total current-current correlation function in \cref{eq:HF} consists of three parts. The first contribution contains the auto-correlation function of the collective mode \(\mathcal{C}_{VV}\), second the cross-correlation function between the collective mode contribution and the Langevin sources \(\mathcal{C}_{IV}\) and lastly the auto-correlation function of the Langevin sources \(\mathcal{C}_{II}\). We find the remarkable result

\begin{gather*}
	\frac{1}{R_q}\int \frac{\diff k}{2 \pi} \mathcal{C}_{VV}= -\int \frac{ \diff k}{2\pi}\left( \mathcal{C}_{IV} + \mathcal{C}_{VI}\right),\\
	\mathcal{C}_{XY}= \left\langle  \delta X^{\alpha'}(k,\omega)\delta Y^{\alpha'}(-k,-\omega)\right\rangle,
\end{gather*}
which holds for the charged and neutral mode, \(\alpha'=c,n\) and \(X,Y\in\{ V,  I\}\). The exact cancellation after taking the momentum integral implies that the Joule heating on the nodes, given by \(\mathcal{C}_{VV}\), is fully compensated by the back action of the sources on the nodes, \(\left( \mathcal{C}_{IV} + \mathcal{C}_{VI}\right)\). The correlation function of the source   \(\mathcal{C}_{II}\) is the only remaining part of \cref{eq:HF}, which explains why there is always a flux quantum for the charged and neutral modes. The sources are thus not just auxiliary but real physical entities of the system. This principle of exact cancellation due to the back action of thermal noise was first mentioned in \cite{nyquist_thermal_1928} by Nyquist who stated that two resistors in thermal equilibrium connected by ideal wires excite thermal fluctuations which in principle leads to a heat flux from one resistor to another, but is compensated by the fluctuations of the second resistor. This also holds true in every frequency window, since one would be able to extract energy by placing a frequency filter in the system. The same is true here. The cancellation holds before the integration over frequencies is done. The simple reason to explain the exact cancellation is the second law of thermodynamics, which forbids to extract heat, i.e. dissipate heat in the present case, if the system is in thermal equilibrium. This is a key difference to \cite{goremykina_heat_2019}, where the energy flux was defined as potential energy flux proportional to \(\mathcal{C}_{VV}\), a contribution which is now canceled by the back action effect of the sources. The implications of this will be discussed in the next section.

\section{Low-energy theory of an edge with intrinsic dissipation}

In this section we want to address the differences between the hydrodynamic model \cite{goremykina_heat_2019} and the transmission line approach presented in this paper and why the former yields different results, despite correctly applying fluctuation-dissipation relations, a standard procedure, to obtain equilibrium correlation functions.

 The general idea is now to obtain a low energy theory from the discrete transmission line model, to compare it to the one obtained in the previous paper  and to comment on its universality. It is clear that in the low-energy limit, e.g. for small temperatures not all possible modes of the non-linear spectrum of the charged and neutral mode will be excited. This justifies to linearize the spectrum, if possible, to the point where the heat flux integrals \cref{eq:hfc,eq:hfn} converge and yield the same heat flux quantum. This is equivalent to finding the low-energy field theory, which correctly describes chiral heat transport in the QH edge in the presence of dissipation. We will discuss the charged and neutral mode separately.

 \subsection{Low-energy field theory for the charged mode}
 
 The dispersion relation of the charged mode is given by \cref{eq:dispc}. It contains a 'hidden' type of dispersion coming from the discreteness of the transmission line. This can be seen, by expanding the dispersion relation in small \(\xi\).

 \begin{equation}\label{eq:specc}
	-i \omega^c(k)\approx -i k v_q -\frac{k^2 \xi v_q}{2}+i\frac{k^3 \xi^2 v_q}{6},
 \end{equation} where we rescaled to intensive quantities by introducing the velocity \(v_q=\xi/R_q C\). The second term plays the role of a dissipative term coming from the retardation of the collective mode and vanishes in the true continuum limit \(\xi \rightarrow0\). Notice that this crossover is non-trivial; physically this means, that the decay length of the collective mode becomes much larger than the distance between the nodes, so no dissipation is happening and thus no heating and back action effect, as discussed in the previous section. In this limit the system is susceptible to it's boundary conditions and the potential energy flux described in \cite{goremykina_heat_2019} correctly predicts a flux quantum. If one keeps dissipation in the system, the definition of the heat flux in terms of the current current correlation function correctly gives a heat flux quantum, if one restricts the energies to be small; see the supplemental material  for details on the calculation. This allows us to write the equation of motion in a coarse grained fashion, i.e. depending on a continuous variable, rather than a discrete node index. This can be understood as the minimal expansion of the discrete difference operator in Kirchhoff's law in order to capture the feature of dissipation correctly at low energies plus a Langevin equation. In real space the equation of motion for the coarse grained collective mode \(V^c(x,t)\) and source \( \delta I^{c}(x,t)\) read
 
 \begin{gather}\label{eq:HD1}
 	\Delta I^c_{\text{out}}(x,t) = \frac{1}{R_q} \delta V^c(x,t) +  \delta I^{c}(x,t),\\\label{eq:HD2}
 	\partial_t \delta V^c(x,t) + v_q R_q  \mathcal{D}  \Delta I^c_{\text{out}}(x,t) = 0
 \end{gather}where  \(\mathcal{D}= \partial_x -\frac{ \xi}{2} \partial_x^2 + \frac{ \xi^2}{6} \partial_x^3\).  With these modified hydrodynamic equations one can proceed as follows: (i) Solve \cref{eq:HD1,eq:HD2} for the collective mode contribution \(\delta V^c\). (ii) The heat flux carried by the chiral system is given by \(J= R_q \left\langle  \left( \Delta I^c_{\text{out}}(x,t)\right)^2 \right\rangle\). (iii) Take into account only low energies and momenta, i.e. expand the cubic and quartic terms in the  momenta around \(k\rightarrow \omega/v_q\), which will result in a heat flux quantum, shown explicitly in the supplemental material. This completes the low-energy hydrodynamic theory for chiral dissipative systems.

\subsection{Low-energy field theory for the neutral mode}

The dispersion relation of the neutral mode is given by \cref{eq:dispn} and contains two types of dissipation. The dissipation due to the discreteness of the transmission line, similarly to the charged mode and the dissipation introduced by the transverse resistor. Focusing on the transverse dissipation only, we keep the leading order terms

\begin{equation}
	 -i \omega^n(k)\approx  -i v_q  k -\frac{v_q\Gamma}{ \xi}.
\end{equation} As we show explicitly in the supplemental material, if \(\Gamma \approx k \xi \ll 1\) we find that the heat integral gives a flux quantum. If the above condition is not met, the integral does not come from low momenta, but from the bandwidth of the system and one needs to integrate over all non-linearities of the spectrum. Physically this means that the neutral collective mode decays over distances shorter than the nodes for stronger dissipations and is essentially pinned. There is no low-energy description of the strongly overdamped neutral mode. The thermal fluctuations however still carry a flux quantum. We are able to write a coarse grained equation of motion for the case that the low energy theory exists   \footnote{One can proceed to solve the equations and complete the heat flux in the same way as described for the charged mode.}.

 \begin{gather}\label{eq:HD3}
	\Delta I^n_{\text{out}}(x,t) = \frac{1}{R_q} \delta V^n(x,t) +  \delta I^{n}(x,t),\\\label{eq:HD4}
	\Delta I^\perp(x,t) = \frac{2}{R_\perp} \delta V^n(x,t) +  \delta I^{\perp}(x,t),\\\label{eq:HD5}
	\partial_t \delta V^n(x,t) + v_q R_q  \partial_x \Delta I^n_{\text{out}}(x,t)-\Delta I^\perp(x,t) = 0.
\end{gather}
We are now able to discuss the role of the artificial cut-off \(\xi_\text{a}\) introduced in \cite{goremykina_heat_2019} for the momentum integration, roughly given by the transverse size of the compressible strip. This is was restriction of the low-energy approach to the edge. We can see that if the integral comes from small energies and momenta, replacing the artificial cut-off by the true cut-off can only change the numerical prefactor, but not the parametrical suppression found in the heat flux of the neutral mode. We want to emphasize once more that the definition of heat flux as potential energy flux was the cause for this result, however the mistake is rather conceptual. It can be seen that the Langevin sources introduced in the present formalism are not just a mathematical trick, but a real part of the system leading to a cancellation of the auto-correlator of the collective mode. The Joule heating is canceled by the back action of the sources on the collective mode in thermal equilibrium. We thus conclude that a Langevin equation is the more complete approach to dissipative systems and we naturally give a more correct definition of energy flux in these systems, which is unobtainable starting from the hydrodynamic point of view.

\section{The role of Aleiner-Glazman modes}

So far we were able to write down an effective model for a system with either two co-propagating modes, i.e. filling factor \(\nu=2\) with a transverse spin resistance or, equivalently after rescaling \(R_q\rightarrow R_q/2 \), the hydrodynamic model consisting of one charged and one half filled Aleiner-Glazman neutral mode, stemming from approximating the density profile obtained by the electrostatic analysis of the edge \cite{aleiner_novel_1994}.

\begin{figure}[htbp] 
	\centering
	\includegraphics[width=0.9\linewidth]{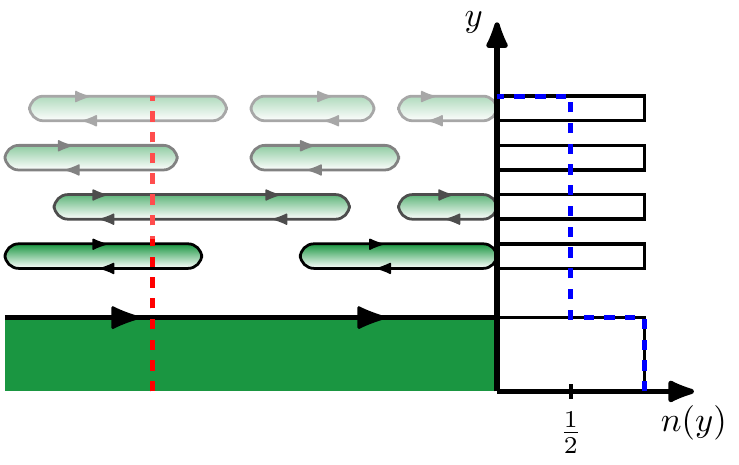}
	\caption{This figure shows an effective description of edge defects as quantum hall puddles and the corresponding density profile in black, cutting through the red dashed line. The blue dashed line shows the ``average" hydrodynamic density, which suggests the presence of a half filled neutral mode.  The model presented in this paper suggests the presence of the disorder length scale in the hydrodynamic equations of motions. This implies that viewing the non-homogenous density profile at the edge due to quantum puddles as an on average ``half filled" neutral mode is wrong, or, if it exists, the neutral mode will decay on length scales comparable to the disorder length scale.}
	\label{fig:alg}
\end{figure}

However writing down an effective model for dissipation in these systems, we see that it is not permitted to write a low energy theory for the neutral mode at all in the presence of dissipation. We are not able to properly construct the continuum limit of the TL based on the edge reconstruction of the ALG modes. Furthermore, the dissipation in the charged mode of the system depends on length scales, e.g. the correlation length of disorder, which are not captured in the low-energy approach. We thus conclude that the electrostatic/hydrodynamic picture is not applicable to describe dissipative chiral transport in Quantum Hall systems and furthermore neglects the experimental truth in the QH edge. The picture of QH puddles, see \cref{fig:alg}, is not compatible with the prediction of ALG modes. The half filling is an artifact of coarse graining many puddles which have co-and counter-propagating modes. This is different from a half-filled co-propagating mode. 

\section{Conclusion}

We have shown that the heat flux quantization in chiral systems in thermal equilibrium is  robust and reflects fundamental thermodynamic laws. We analyzed chiral quantum systems with the 'transmission line approach', an extension of the Langevin formalism in combination with scattering states developed in earlier works \cite{sukhorukov_scattering_2016,slobodeniuk_equilibration_2013}. The formalism allows to take into account dissipation on an effective level; Similarly to the Caldeira-Leggett model, we attach a bath consisting of many oscillators to a chiral system. In contrast to the Caldeira-Leggett model, we consider the baths to be an open system, which allows for the analytical treatment of equilibrium and non-equilibrium situations. We were not only able to address the question of the apparent reduction of heat flux carried by the strongly overdamped neutral mode in \cite{goremykina_heat_2019}, but have proven the quantization of heat flux for chiral systems in the presence of different types of dissipation. In a next step we considered  chiral systems in the presence of additional diffusive modes, making the system essentially non-chiral.

For all of the systems above we could proof that heat flux is quantized if the system is in thermal equilibrium and could relate this to the local unitarity condition of the scattering matrix and the back action principle of the Langevin sources on the collective mode. We have shown the correct definition of heat flux for chiral dissipative systems and discussed the crossover to the potential energy flux used in \cite{goremykina_heat_2019}. This question of the definition of energy flux reveals the nature of the Langevin sources as true physical entities which are able to give back heat to the system rather than just being a mathematical aid. We continued formulating the low-energy theories which can be deduced from the transmission line model and capture the universal aspects of the edge. We discussed the minimal expansion of Kirchhoff's law, which completes the low-energy theory.

The results of this paper are not able to answer the missing heat paradox \cite{le_sueur_energy_2010}, but open a new field containing many new ideas. An explanation for the experiment could be a potential discrepancy between the actual and measured heat flux. To proceed further one needs to look into the physics of probing the edge states. This involves the construction of a new electronic operator which is not just the bosonized vertex operator but goes beyond the bosonization formalism due to correlations through the collective mode if a tunnel probe is connected. The results of this paper also shows that the strongly overdamped AG modes cannot be described in a low-energy theory and if they are present cannot explain the experiment. So far their experimental detection remains an open question. Finally, the formalism developed in this paper can be easily applied to different edge reconstructions and to FQH systems.

The authors acknowledges the financial support from
the Swiss National Science Foundation.

\bibliographystyle{apsrev4-2}
\bibliography{TL_Paper.bib}
\cleardoublepage
\onecolumngrid
\appendix

\begin{center}
	{\large
		{\bf
			Supplemental Materials for the manuscript \\ \vspace{0.25cm}
		}
	``Transmission line approach to transport of heat in chiral systems with dissipation" \\ \vspace{0.25cm}
	}
	by F. St{\"a}bler and E. Sukhorukov
\end{center}

\section{Solution of the equation of motion and computation of the heat flux}\label{app:eom}

We use the following convention for the Fourier transformation

\begin{equation} 
	X_j(t)=\sum\limits_{k=0}^{N-1} \int \frac{\diff \omega}{2\pi} e^{i\frac{2 \pi k}{N} j-i\omega t} X_k(\omega),\qquad 	X_k(\omega)=\frac{1}{N}\sum\limits_{j=0}^{N-1} \int \diff t  e^{-i \frac{2 \pi k}{N}  j+i \omega t} X_j(t).
\end{equation}

The solution of the equation of motion \cref{eq:VC,eq:VN} after the Fourier transformation are given by

\begin{equation}
	\delta V_k^c(\omega) = \frac{  \delta I_k^{c}(\omega) R_q \omega_k^c}{\omega-\omega_k^c},\qquad
	\delta V_k^n(\omega) = \frac{  \delta I_k^{n}(\omega) R_q \omega_k^c- i \delta I_\perp/C}{\omega-\omega_k^n}.
\end{equation} where  the spectrum of the charged an neutral modes can be read off

\begin{equation}
		-i \omega_k^c=\left(R_q C\right)^{-1}\left(e^{-2 \pi i k /N}-1\right), \qquad 	-i \omega_k^n=\left(R_q C\right)^{-1}\left(e^{-2 \pi i k /N}-1-\Gamma\right).
\end{equation}
 The heat flux in the upper and lower arm is given by the outgoing current-current correlation function.

\begin{equation}
	J = \frac{\pi \hbar}{e^2} \left\langle \left(\Delta I^{(1)}_{\text{out},j}\right)^2 +\left(\Delta I^{(2)}_{\text{out},j}\right)^2 \right\rangle =R_q \left\langle \left(\Delta I^{c}_{\text{out},j}\right)^2 +\left(\Delta I^{n}_{\text{out},j}\right)^2 \right\rangle.
\end{equation} We  compute the contributions of to the heat flux of the charged mode \(J^c\) and the neutral mode \(J^n\) separately.

\subsection{Heat flux of the charged mode}

For the charged mode we find

\begin{multline}\label{eq:JCdirect}
	J^c=R_q \left\langle \left(\Delta I^{c}_{\text{out},j}\right)^2  \right\rangle = R_q\sum_{k,q} \int \frac{\diff\omega \diff \omega'}{4\pi^2}  e^{i\frac{2 \pi (k+q)}{N} j-i(\omega+\omega') t}  \left\langle \Delta I^{c}_{out,k}(\omega)\Delta I^{c}_{out,q}(\omega') \right\rangle \\ =R_q \int \frac{\diff\omega \diff k}{4\pi^2} \frac{ R_q^2 C^2 \omega^2  S_c(\omega)}{2+R_q^2 C^2 \omega^2 -2 \cos(k)- 2 R_q C \omega \sin(k)},
\end{multline} where we took the large N limit \(\frac{1}{N}\sum_{k=0}^{N-1} \rightarrow \int_{0}^{2\pi} \frac{\diff k}{2\pi} \). The sources are delta-correlated \(\left\langle \delta I^{c}_{out,k}(\omega)\delta I^{c}_{out,q}(\omega') \right\rangle = \frac{\delta_{k,-q}}{N}2\pi \delta(\omega+\omega') S_c(\omega)\), with \(S_c(\omega)= \frac{1}{4}(S_1+S_2) =\frac{1}{2 R_q} \left(\frac{\hbar \omega}{1-e^{-\beta \hbar \omega}}-\hbar \omega \theta(\omega)\right)\).

The \(k\)-integral can be mapped onto the unit circle by the transformation \(z=e^{i k}\) or equivalently \(k=-i \log(z)\) (\(\diff k = \diff z/(iz)\)). The integral has one pole inside the unit circle at \(z^*=(1-i R_q C \omega)^{-1}\). We thus find

\begin{multline}
	J^c= R_q \int \frac{\diff\omega \diff z}{4\pi^2} \frac{ i R_q^2 C^2 \omega^2  S_c(\omega)}{1+i R_q C \omega-z(2+R_q^2 C^2 \omega^2)+z^2(1-i R_q C \omega)}=R_q \int \frac{\diff\omega}{2\pi} S_c(\omega)\\=  \int \frac{\diff\omega}{4\pi} \left[\frac{\hbar \omega}{1-e^{-\beta \hbar \omega}}-\hbar \omega \theta(\omega)\right]=\frac{1}{4\pi} \frac{\pi^2}{3 \hbar \beta^2}=\frac{\pi}{12 \hbar \beta^2} = J_q.
\end{multline}

\subsection{Heat flux of the neutral mode}

We repeat the same procedure for the neutral mode \(J^n\)

\begin{multline}
		J^n= R_q \left\langle \left(\Delta I^{n}_{\text{out},j}\right)^2  \right\rangle = R_q \sum_{k,q} \int \frac{\diff\omega \diff \omega'}{4\pi^2}  e^{i\frac{2 \pi (k+q)}{N} j-i(\omega+\omega') t}  \left\langle \Delta I^{n}_{out,k}(\omega)\Delta I^{n}_{out,q}(\omega') \right\rangle\\= R_q \int \frac{\diff\omega \diff k}{4\pi^2} \frac{ R_q^2(1+R_\perp^2 C^2 \omega^2)  S_n(\omega)+R_\perp^2  S_\perp(\omega)}{2 R_q R_\perp+2
			R_\perp^2+R_q^2 \left(1+C^2 R_\perp^2 \omega ^2\right)-2 R_\perp \left(  \left(R_q+R_\perp\right)  \cos  \left(k\right)   + R_\perp       R_q C  \omega  \sin \left(k\right) \right)},
\end{multline} which can be rewritten in using the \(z\) variable. The integral has one pole inside the contour at \(z^*=(1+\frac{RQ}{R_\perp}-i R_q C \omega)^{-1}\). This gives

\begin{multline}
	J^n= R_q \int \frac{\diff\omega \diff z}{4\pi^2} \frac{ i\left( R_q^2\left(1+R_\perp^2 C^2 \omega^2\right)  S_n(\omega) +R_\perp^2 S_\perp(\omega)\right)}{\left(R_q+R_\perp (1-z)+ i R_\perp R_q C \omega\right)\left(R_q z+ R_\perp (z-1)-i R_\perp R_q C \omega\right)}\\= R_q \int \frac{\diff\omega}{2\pi} \frac{R_q^2\left(1+R_\perp^2 C^2 \omega^2 \right)S_n(\omega)+R_\perp^2 S_\perp(\omega)}{R_q(R_q+R_q R_\perp^2 C^2\omega^2+2R_\perp)}=  \int \frac{\diff\omega}{4\pi} \left[\frac{\hbar \omega}{1-e^{-\beta \hbar \omega}}-\hbar \omega \theta(\omega)\right]=J_q,
\end{multline} where \(S_n(\omega)= \frac{1}{4}(S_u+S_d) =\frac{1}{2 R_q} \left(\frac{\hbar \omega}{1-e^{-\beta \hbar \omega}}-\hbar \omega \theta(\omega)\right)\) and \(S_\perp(\omega)= \frac{2}{R_\perp} \left(\frac{\hbar \omega}{1-e^{-\beta \hbar \omega}}-\hbar \omega \theta(\omega)\right)\).

\subsection{Cancellation of auto and cross correlation functions}

We point out that there is a general cancellation of collective mode auto and cross correlation function. This can be seen from evaluating the correlation functions appearing in \cref{eq:JCdirect} explicitly, i.e. we find the auto and crosscorrelations of the collective charged mode and sources individually.

\begin{multline}
	\frac{1}{R_q^2}\int \frac{\diff\omega \diff k}{4\pi^2} \left\langle  \delta V^{c}(k,\omega)\delta V^{c}(-k,-\omega)\right\rangle \\=  \int \frac{\diff\omega \diff k}{4\pi^2} \frac{ 2\left(1- \cos \left(k \right)\right)  S_c(\omega)}{2+R_q^2 C^2 \omega^2 -2 \cos(k)- 2 R_q C \omega \sin(k)}= \int \frac{\diff\omega}{2\pi} \frac{2 S_c(\omega)}{1+R_q^2C^2\omega^2},
\end{multline}and 

\begin{multline}
	\frac{1}{R_q}\int \frac{\diff\omega \diff k}{4\pi^2} \left\langle  \delta V^{c}(k,\omega)\delta I^{\parallel,c}(-k,-\omega)+ \delta I^{\parallel,c}(k,\omega)\delta V^{c}(-k,-\omega)\right\rangle \\=  \int \frac{\diff\omega \diff k}{4\pi^2} \frac{2 \left( R_q C \omega  \sin \left(k\right)-2\left(1- \cos (k)\right)\right)  S_c(\omega)}{2+R_q^2 C^2 \omega^2 -2 \cos(k)- 2 R_q C \omega \sin(k)}= -\int \frac{\diff\omega}{2\pi} \frac{2 S_c(\omega)}{1+R_q^2C^2\omega^2},
\end{multline}

where we used the same methods as before to calculate the momentum integral. The same can be shown for the neutral mode.

\section{IR-theory from exact theory}

In real experimental situations the temperature is usually small and only the low-energy part of the spectrum will be excited. We will expand the dispersion relation up to the point where the heat integrals converge giving a correct low energy description for the chiral edge. All nonlinear contributions will be small in the continuum limit, but are nonetheless important for the understanding of the low-energy theory. This allows us to do three simplifications: (i) The main contribution of the integral comes from small momenta. The cut-off can be taken to infinity. We will estimate the error of the tails of the integral at a later point. (ii) The parameter \( \xi\) is small in the continuum limit and can be treated pertubatively. (iii) We rescale the capacitance of the system, introducing the velocity \(v_q = \frac{\xi}{R_q C}\). We assume that only the low-energy part of the spectrum is excited, i.e. \(\omega \approx v_q k\) and we can introduce a small parameter \(\tilde{k}=\omega - v_q k\), which we can treat perturbatively.

\subsection{Low energy theory of the charged mode}\label{app:FTc}

We restore the distance between the nodes \(\xi\). The heat flux is given by

\begin{multline}
	J^c= \frac{1}{8\pi^2} \int \diff \omega \int_{-\pi/\xi}^{\pi/\xi} \diff k  \frac{ \xi \omega^2  }{\left(\omega+\omega^c(k)\right)\left(\omega-\omega^c(k)\right)} S_0(\omega)\\=\frac{1}{8\pi^2} \int \diff \omega \int_{-\pi/\xi}^{\pi/\xi} \diff k  \frac{ \xi R_q^2 C^2 \omega^2  }{\left(R_q C \omega-k \xi\right)^2+\frac{1}{3} k^3 \xi^3 R_q C \omega -\frac{1}{12} k^4 \xi^4  }S_0(\omega)\\=\frac{1}{8\pi^2} \int \diff \omega \int_{-\pi/\xi}^{\pi/\xi} \diff k  \frac{ \xi \omega^2  }{\left(\omega- v_q k \right)^2+\frac{1}{3} k^3 \xi^2 v_q \omega -\frac{1}{12} k^4 \xi^2 v_q^2  }S_0(\omega), 
\end{multline} where we expanded the dispersion up to third order using the smallness of  \(\xi\), which gives \(	\omega^c(k)\approx k v_q -i\frac{k^2 \xi v_q}{2}-\frac{k^3 \xi^2 v_q}{6},\) and \(S_0(\omega)=\frac{\hbar \omega}{1-e^{-\beta \hbar \omega}}-\hbar \omega \theta(-\omega)\). For \(\xi \rightarrow 0\) the integrand has a singularity  \(k^* = \omega/v_q\). Deviations from the singularity  are small for  small \(\xi\), as required by the low energy approach. We can thus shift the integral by \(\tilde{k} \rightarrow k - \omega/v_q\) and expand the cubic and quartic part to order \(\mathcal{O}(\tilde{k}^0)\).

\begin{multline}
	J^c= \frac{1}{8\pi^2} \int \diff \omega \int_{-\pi/\xi}^{\pi/\xi} \diff k  \frac{\xi \omega^2  }{\left(\omega- v_q k \right)^2+\frac{1}{3} k^3 \xi^2 v_q \omega -\frac{1}{12} k^4 \xi^2 v_q^2  }S_0(\omega) \\ \sim \frac{1}{8\pi^2} \int \diff \omega \int_{-\infty}^{\infty} \diff \tilde{k}  \frac{ \xi \omega^2  }{ v_q^2 \tilde{k}^2 + \frac{1}{4}\frac{\omega^4 \xi^2}{v_q^2} }S_0(\omega) =  \int  \frac{\diff \omega}{4\pi} S_0(\omega) = J_q.
\end{multline}

%
%
%
%
%
%

\subsection{Low energy theory of the neutral mode}\label{app:FTn}

The spectrum of the neutral mode is different. It contains now also the transverse resistor.  To have a meaningful low-energy theory we want to analyze the heat flux in the limit of infinitesimal \(\Gamma = R_q/R_\perp\). This additional term regularizes the integral already in the order \(\mathcal{O}(k^2)\) of the dispersion relation. The heat flux of the neutral mode is given by

\begin{multline} 
 J^n=\frac{1}{4\pi^2} \int \diff \omega \int_{-\infty}^{\infty} \diff k \xi \frac{ S_n(\omega) R_q^2  (1+R_\perp^2 C^2 \omega^2)+R_\perp^2 S_\perp(\omega) }{R_q\left(1+i R_\perp C \omega - R_\perp C \lambda(-k)\right)\left(1-i R_\perp C \omega - R_\perp C \lambda(k)\right)} \\= \frac{R_q}{4\pi^2} \int \diff \omega \int_{-\infty}^{\infty} \diff k \xi \frac{ S_n(\omega)   (\Gamma^2+R_q^2 C^2 \omega^2)+ S_\perp(\omega) }{(R_q C \omega - k\xi)^2 +\Gamma^2+k^2 \xi^2 \Gamma} \\= \frac{1}{4\pi^2} \int \diff \omega \int_{-\infty}^{\infty} \diff k \xi \frac{ \frac{1}{2}   (\Gamma^2+R_q^2 C^2 \omega^2)+ \Gamma}{(R_q C \omega - k\xi)^2 +\Gamma^2+k^2 \xi^2 \Gamma}S_0(\omega)\\\sim \frac{1}{4\pi^2} \int \diff \omega \int_{-\infty}^{\infty} \diff k  \frac{ \frac{\Gamma}{\xi}}{( \frac{\omega}{v_q} - k)^2 +\frac{\Gamma^2}{\xi^2}}S_0(\omega)
 = J_q,\label{eq:LENeutral}
\end{multline} we use the fact that \(\Gamma \approx k \xi \ll 1 \). This means that terms like \(\Gamma \frac{\Gamma}{\xi}\), \(k^2 \Gamma \) and \(\omega^2/v_q^2 \xi  \) will be of next leading order. The same integral as the last line of \cref{eq:LENeutral} can be also obtained by only taking into account the terms $  \omega^n(k)\approx   v_q  k -i\frac{v_q\Gamma}{ \xi}.$ of the spectrum. Thus the integral is regularized already at the linear order of the expansion of the spectrum.

\end{document}